\documentclass[iicol,pdflatex,sn-mathphys-num]{sn-jnl}

\usepackage{stfloats}
\usepackage{cuted}
\usepackage{graphicx}%
\usepackage{multirow}%
\usepackage{amsmath,amssymb,amsfonts}%
\usepackage{amsthm}%
\usepackage{mathrsfs}%
\usepackage[title]{appendix}%
\usepackage{xcolor}%
\usepackage{textcomp}%
\usepackage{manyfoot}%
\usepackage{booktabs}%
\usepackage{algorithm}%
\usepackage{algorithmicx}%
\usepackage{algpseudocode}%
\usepackage{listings}%
\usepackage{svg}
\usepackage{booktabs}
\usepackage{threeparttable}

\theoremstyle{thmstyleone}%
\theoremstyle{thmstyletwo}%

\theoremstyle{thmstylethree}%

\begin{document}

\title[Article Title]{Measuring trainable degrees of freedom in materials graph neural networks: a random-subspace intrinsic dimension analysis}

\author[1,2]{Shehroz Ahmad Shoaib}

\author*[1]{Kangming Li}\email{kangming.li@kaust.edu.sa}

\affil*[1]{Physical Science and Engineering Division, King Abdullah University of Science and Technology, Thuwal, Saudi Arabia}

\affil[2]{Department of Electrical Engineering, King Fahd University of Petroleum and Minerals, Dhahran, Saudi Arabia}

\abstract{
Final predictive accuracy is the standard basis for comparing graph neural networks (GNNs) in materials-property prediction, but it does not show how strongly performance depends on access to trainable parameter-space directions. Here, we introduce trainable-degree dependence as a complementary characterization of materials GNN learning. Using random-subspace intrinsic-dimension analysis, we train CGCNN, ALIGNN, and DimeNet++ in randomly oriented parameter subspaces across six prediction tasks and measure how performance recovers as independent trainable degrees of freedom are restored. The resulting recovery curves separate endpoint accuracy from the trainable-dimensional demand required to recover it. They reveal distinctions that final errors alone miss: metallic classification and log-bulk-modulus regression recover near-reference performance from small fractional subspaces, formation-energy and band-gap prediction show stronger architecture dependence, and phonon prediction is most sensitive to dimensional restriction. Dataset-size sweeps show that band-gap models require larger fractional subspaces as training data grows, whereas formation-energy and bulk-modulus responses are more stable. A width sweep shows that fractional thresholds can remain stable while absolute threshold dimensions increase with model size. Random-subspace analysis therefore provides a targeted stress test for how materials GNNs use their optimization space.

}

\maketitle
\section*{Introduction}\label{sec1}

Atomistic graph neural networks (GNNs) have become a central tool for predicting materials properties~\cite{reiser2022graph,choudhary2022recent,merchant2023scaling}. Advances in atomistic GNN architectures have shown that structural representations can learn mappings from atomic arrangements to properties including band gap, formation energy, elastic moduli, dielectric response, and vibrational quantities~\cite{xie2018cgcnn,park2020developing,gasteiger2020dimenetplusplus,choudhary2021alignn,yan2022periodic,10.1039/d4dd00018h,dunn2020matbench,choudhary2024jarvis,riebesell2023matbenchdiscovery,fung2021benchmarking}. Model comparisons nevertheless remain dominated by final predictive accuracy, commonly reported as mean absolute error (MAE). Although indispensable, aggregate accuracy alone does not reveal whether performance arises from a favorable architectural inductive bias, dominant chemical trends, or the composition and coverage of the training data~\cite{gasteiger2020dimenetplusplus,choudhary2021alignn,wang2021compositionally,li2023exploiting}. It also provides little information about how strongly the learned solution depends on access to the model's full optimization space~\cite{li2018intrinsicdimension,ding2023parameter}.

This limitation has practical consequences for materials dataset design. Increasing dataset size may introduce new chemistries, symmetries, and bonding motifs, but it may also increase the sampling density of regions that are already well represented~\cite{li2023critical,li2023exploiting,zhang2023entropy,li2024md,qi2024robust}. Conversely, reducing a dataset can simplify the apparent learning problem while removing examples needed for reliable generalization~\cite{jain2013materialsproject,dunn2020matbench,omee2024ood}. Dataset growth may therefore affect not only attainable accuracy, but also the complexity and reproducibility of the learned mapping.

A related issue arises in model design. Materials GNNs encode different structural assumptions: CGCNN~\cite{xie2018cgcnn} applies graph convolutions over atoms and bond features, DimeNet++~\cite{gasteiger2020dimenetplusplus} incorporates directional information through angular message passing, and ALIGNN~\cite{choudhary2021alignn} uses a line-graph representation to capture bond-angle information. Such architectural choices may be advantageous for different target properties, while poor performance may reflect incomplete structural representations, limited data, label noise, or mismatch between the architecture and the target~\cite{bartel2020critical,dunn2020matbench,ottomano2024not,omee2024ood}. A model that is well suited to formation-energy prediction may therefore not learn phonon-related quantities with the same efficiency. Final MAE alone does not show how much of the available parameter space each model requires to attain its performance.

Random-subspace analysis provides a complementary way to ask how a model uses its trainable parameter space. Conventional model-scaling studies alter the architecture and examine performance as a function of the native parameter count $D$~\cite{nakkiran2021deep,frey2023neural}, while pruning methods remove selected weights or structured components~\cite{lecun1990obd,han2015learning,liu2017learning}. Random-subspace analysis instead holds the $D$-parameter architecture fixed and restricts optimization to a randomly oriented coordinate system of dimension $d\leq D$~\cite{li2018intrinsicdimension}. Sweeping $d$ therefore does not test whether the model can be compressed; it tests how the model's predictive performance recovers as independent trainable directions are progressively restored.

This recovery curve separates quantities that are usually conflated in a single benchmark score. The full-subspace endpoint measures the attainable predictive accuracy of the model under the chosen training protocol, whereas the shape of the curve measures how strongly that accuracy depends on the number of trainable degrees of freedom. A model--task pair that reaches low error only when a large fraction of the training space is available is therefore different from one that reaches similar error with a much smaller fraction, even if their final MAEs are comparable. The same curve also exposes run-to-run variability, providing a measure of how reproducibly near-reference performance is recovered under restricted training freedom.

Here, we apply this analysis to materials-property prediction using CGCNN, ALIGNN, and DimeNet++ across six regression and classification tasks. For each model--task combination, we measure how predictive performance changes as the trainable subspace dimension is increased, and we summarize this behavior using both the full recovery curve and the threshold dimension required to recover near-reference performance. We then examine how these responses vary with target property, dataset size, model architecture, and model width, and we test their sensitivity to the construction of the random projection.

The analysis shows that trainable-degree demand is an informative, model-relative property of a materials-learning system. It is not determined by final predictive accuracy alone: different properties exhibit distinct recovery curves, architectures can trade off absolute accuracy against dimensional demand, dataset growth can change either the mean response or its stability across runs, and model-width scaling can preserve fractional thresholds while increasing the absolute number of required trainable dimensions. These findings position random-subspace intrinsic-dimension analysis as a complementary diagnostic for materials GNNs. It does not identify a compressed model or assign an architecture-independent physical complexity to a materials property; rather, it reveals how a particular model, dataset, and target property interact through the available optimization space.

\section*{Results}\label{sec3}

\subsection*{Intrinsic-dimension analysis}

\begin{figure}[tb]
\centering
\includegraphics[width=\linewidth]{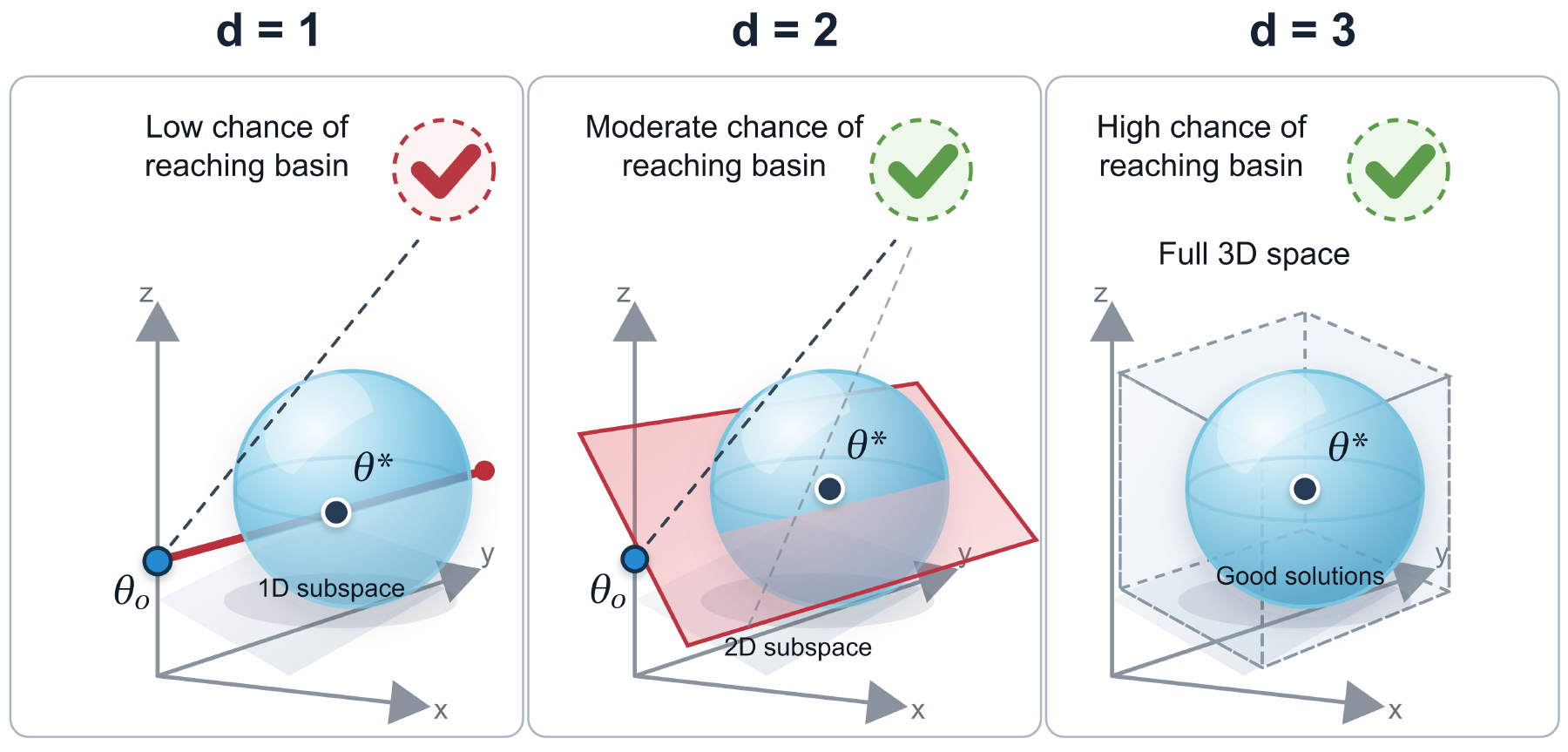}
\caption{Schematic of random-subspace training in a toy parameter space with ambient dimension $D=3$. A model is initialized at $\boldsymbol{\theta}_0$, and optimization is restricted to an affine subspace $\boldsymbol{\theta}_0+\mathbf{A}\boldsymbol{\phi}$ of dimension $d\leq D$. The blue region represents parameter configurations that satisfy the target performance criterion, with $\boldsymbol{\theta}^{*}$ denoting a representative solution. As $d$ increases, a randomly oriented subspace is more likely to intersect the acceptable-solution region; when $d=D$, the full parameter space is available for optimization.}
\label{fig_intrinsic}
\end{figure}

To quantify how predictive performance depends on the available trainable degrees of freedom, we train each GNN within randomly oriented parameter subspaces of increasing dimension. Let $D$ denote the ambient number of model parameters and $d\leq D$ the dimension of the trainable subspace. For a fixed $d$, optimization is performed over $d$ trainable coordinates whose updates are mapped into the full parameter space through a fixed random projection. The resulting variation in predictive performance with $d$ defines the dimensional-response curve of a model-task combination, as illustrated in Figure~\ref{fig_intrinsic}.

Unless otherwise stated, we present each sweep as a normalized predictive metric plotted against the fractional subspace dimension $d/D$. For regression tasks, the normalized error is defined as
\[
\widetilde{\mathrm{MAE}}(d)=\frac{\mathrm{MAE}(d)}{\mathrm{MAE}(D)},
\]
whereas for the is-metal classification task the normalized score is
\[
\widetilde{\mathrm{AUC}}(d)=
\frac{\mathrm{ROC\mbox{-}AUC}(d)}{\mathrm{ROC\mbox{-}AUC}(D)}.
\]
A value of one therefore corresponds to the performance of the $d=D$ projected run. This normalization enables comparison across properties with different target scales and evaluation metrics.

When a scalar summary is required, we use the threshold dimension $d_{10}$, defined as the dimension required to recover performance within a 10\% relative tolerance of the corresponding $d=D$ reference. This corresponds to a normalized MAE no greater than 1.10 for regression and a normalized ROC-AUC no lower than 0.90 for classification. In the main cross-architecture comparisons, we emphasize the fractional threshold $d_{10}/D$ because the three models have approximately matched ambient dimensions. Absolute dimensions and unnormalized predictive errors are also retained to establish the underlying performance scale. When model size is varied, we report both $d_{10}$ and $d_{10}/D$, because the absolute and fractional measures need not exhibit the same scaling behavior. Each sweep is referenced to its own $d=D$ run obtained using the same projected parameterization, thereby isolating the degradation associated with restricting the trainable subspace. Formal definitions and direct-training controls are provided in the Methods Section.

We apply the analysis to CGCNN, ALIGNN, and DimeNet++ across six tasks: formation energy per atom, band gap, is-metal classification, log bulk modulus, refractive index, and phonon-frequency prediction. To examine the effect of dataset size, we repeat the formation-energy, band-gap, and log-bulk-modulus experiments using the full datasets and randomly sampled one-quarter and one-tenth subsets. Dataset sources, sample sizes, evaluation metrics, and $d=D$ performance are summarized in Table~\ref{tab:full_dim_matbench_comparison}.

Because each dimensional sweep requires repeated training across multiple values of $d$, datasets, and random seeds, we use compact one-layer model configurations and Fastfood structured projections to make the experimental design computationally tractable. The architectures are approximately parameter matched, with ambient dimensions between 83,685 and 85,953. These compact configurations are intended to support controlled comparisons rather than to reproduce the state-of-the-art performance of the original architectures. Nevertheless, their $d=D$ results retain substantial predictive signal: the regression MAEs are below the corresponding dataset MAD values, and the is-metal ROC-AUC values are well above 0.5. Dense and orthonormalized projection variants are evaluated separately in the Supplemental Section (Figure~S1), while complete architecture specifications, optimization hyperparameters, seed definitions, and dataset-sampling procedures are reported in the Methods Section.

\begin{table*}[bt]
\centering
\caption{Dataset sizes and $d=D$ predictive performance of the compact, approximately parameter-matched models. Regression tasks are evaluated using mean absolute error (MAE; lower is better), and the metallicity classification task using ROC-AUC
(higher is better). The dummy reference is the dataset mean absolute deviation for regression tasks and 0.5 for metallicity classification.}
\label{tab:full_dim_matbench_comparison}
\small
\begin{tabular*}{\textwidth}{@{\extracolsep{\fill}}l l r c c c c@{}}
\toprule
Task & Data source & $N$ & Dummy & ALIGNN & DimeNet++ & CGCNN \\
\midrule
Phonon frequency
    & matbench\_phonons
    & 1{,}265 & 323.8 & 73.4 & 87.3 & 108.0 \\

Refractive index
    & matbench\_dielectric
    & 4{,}764 & 0.809 & 0.324 & 0.611 & 0.512 \\

Bulk modulus
    & matbench\_log\_kvrh
    & 110k & 0.290 & 0.064 & 0.087 & 0.071 \\

Band gap
    & MP23
    & 153k & 1.327 & 0.332 & 0.363 & 0.368 \\

Formation energy
    & MP23
    & 155k & 1.059 & 0.048 & 0.060 & 0.053 \\

\addlinespace
Metallicity
    & matbench\_is\_metal
    & 106k & 0.500 & 0.949 & 0.948 & 0.937 \\
\bottomrule
\end{tabular*}
\end{table*}

\subsection*{Different prediction tasks exhibit distinct dimensional signatures}
\begin{figure*}[tb]
\centering
\includegraphics[width=\textwidth]{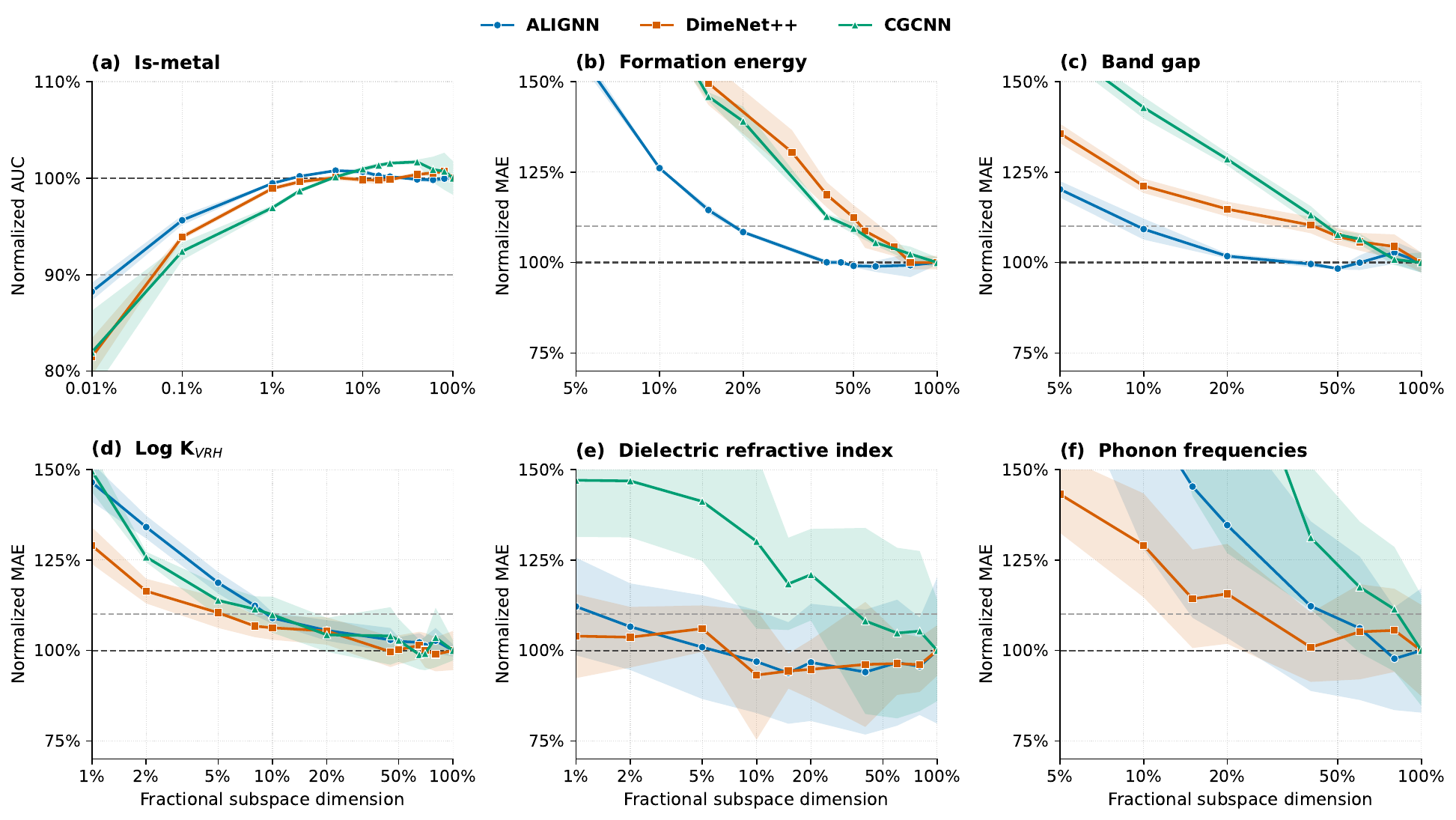}
\caption{Normalized predictive performance as a function of fractional subspace dimension $d/D$ for three GNN architectures across six tasks. Lines and markers denote the mean across repeated runs, and shaded bands show the standard deviation under the randomization protocol described in the Methods Section. The horizontal black line marks the $d=D$ reference, and the gray dashed line marks the 10\% performance threshold used to define $d_{10}$.}
\label{fig_tasks_combined}
\end{figure*}

We first examine how restricting the available trainable subspace affects predictive performance across different materials-property tasks. Figure~\ref{fig_tasks_combined} compares the dimensional responses of CGCNN, ALIGNN, and DimeNet++ across five regression tasks and one classification task. Here, the dimensional signature of a model--task combination refers to three features: the fractional dimension required to recover near-reference performance, the rate at which performance deteriorates below this regime, and the variability across repeated runs. 

The six tasks exhibit qualitatively different signatures. Is-metal classification and log $K_{\mathrm{VRH}}$ recover near-reference performance at relatively small fractional dimensions and show limited separation among architectures. Formation energy and band gap exhibit stronger architecture dependence, whereas dielectric and phonon prediction show greater run-to-run variability, with phonons also displaying the most sustained degradation as the trainable subspace is reduced.

The is-metal classification task exhibits the lowest fractional dimensional demand and the weakest architecture dependence among the tasks considered. The three curves largely overlap and approach their respective $d=D$ ROC-AUC values once approximately 5\% of the ambient dimension is available. At intermediate dimensions, CGCNN slightly exceeds its $d=D$ reference, whereas ALIGNN and DimeNet++ remain close to it. This small improvement is consistent with a mild regularization effect from restricting the optimization space~\cite{nakkiran2021deep}, although its magnitude is modest relative to the overall classification performance. 

Formation energy prediction shows substantially stronger architecture dependence. ALIGNN enters the 10\% tolerance region at 15\% of the ambient dimension, whereas CGCNN and DimeNet++ require roughly one half or more. Because the compact models have closely matched ambient dimensions, this result indicates lower fractional dimensional demand for ALIGNN under the present configurations. Band-gap prediction displays a similar but more sharply separated architecture dependence, especially at smaller fractional dimensions. ALIGNN recovers performance within the 10\% tolerance of its $d=D$ reference using 10\% of the ambient dimension. By comparison, DimeNet++ and CGCNN require substantially larger fractional subspaces, approaching the tolerance region only when roughly half of their parameter spaces are available. 

Similar to the is-metal classification task, the log-$K_{\mathrm{VRH}}$ task shows little separation among the three architectures. All three models recover MAEs within approximately 10\% of their respective $d=D$ references once about 10\% of the ambient dimension is available. Below this regime, their errors increase rapidly as the subspace is further restricted. The similarity of the three curves indicates that, under the compact configurations studied here, log-$K_{\mathrm{VRH}}$ prediction has relatively low fractional dimensional demand and limited sensitivity to the choice among these architectures.

The dielectric task exhibits a qualitatively different response. The mean ALIGNN and DimeNet++ curves approach their $d=D$ performance at relatively small fractional dimensions, whereas the mean CGCNN curve remains above its reference over a larger portion of the sweep. The curves are also nonmonotonic, making a single threshold less representative of the overall response.

Phonon prediction shows the strongest and most sustained degradation as the trainable subspace is reduced. DimeNet++ approaches its $d=D$ reference at a smaller fractional dimension than ALIGNN, whereas CGCNN requires a comparatively large fraction of its parameter space. The low-dimensional values for all three architectures exceed the plotted range, indicating strong sensitivity to dimensional restriction. The normalized phonon responses must also be interpreted together with the absolute errors in Table~\ref{tab:full_dim_matbench_comparison}. At $d=D$, ALIGNN achieves the lowest phonon MAE compared withDimeNet++ and CGCNN (Table~\ref{tab:full_dim_matbench_comparison}). DimeNet++ is therefore less accurate than ALIGNN in absolute terms but shows a less pronounced mean degradation over part of the dimensional sweep. This contrast illustrates an accuracy--dimensional-demand trade-off that is not captured by final MAE alone.

For both dielectric and phonon prediction, however, the uncertainty bands are broad and substantially overlapping over much of the dimensional range. Importantly, variability remains at $d=D$, indicating that the instability is not caused solely by restricting the trainable subspace. These are also the two smallest datasets considered in this study, suggesting that limited sample size may contribute to the observed run-to-run variation by providing weaker constraints on the learned structure--property relationship. Nevertheless, dataset size is not varied independently for these tasks, and the present results cannot distinguish its contribution from those of dataset partitioning, model initialization, projection orientation, and optimization settings. Precise threshold estimates and architecture rankings are therefore less robust for dielectric and phonon prediction than for formation energy, band gap, or log $K_{\mathrm{VRH}}$.

Overall, the normalized dimensional-response curves provide information that is complementary to full-subspace predictive accuracy. They reveal how strongly each model--task combination depends on the available trainable freedom and how reproducible that dependence is across repeated runs.

\subsection*{Dataset size changes dimensional response}

\begin{figure*}[t]
\centering
\includegraphics[width=\textwidth]{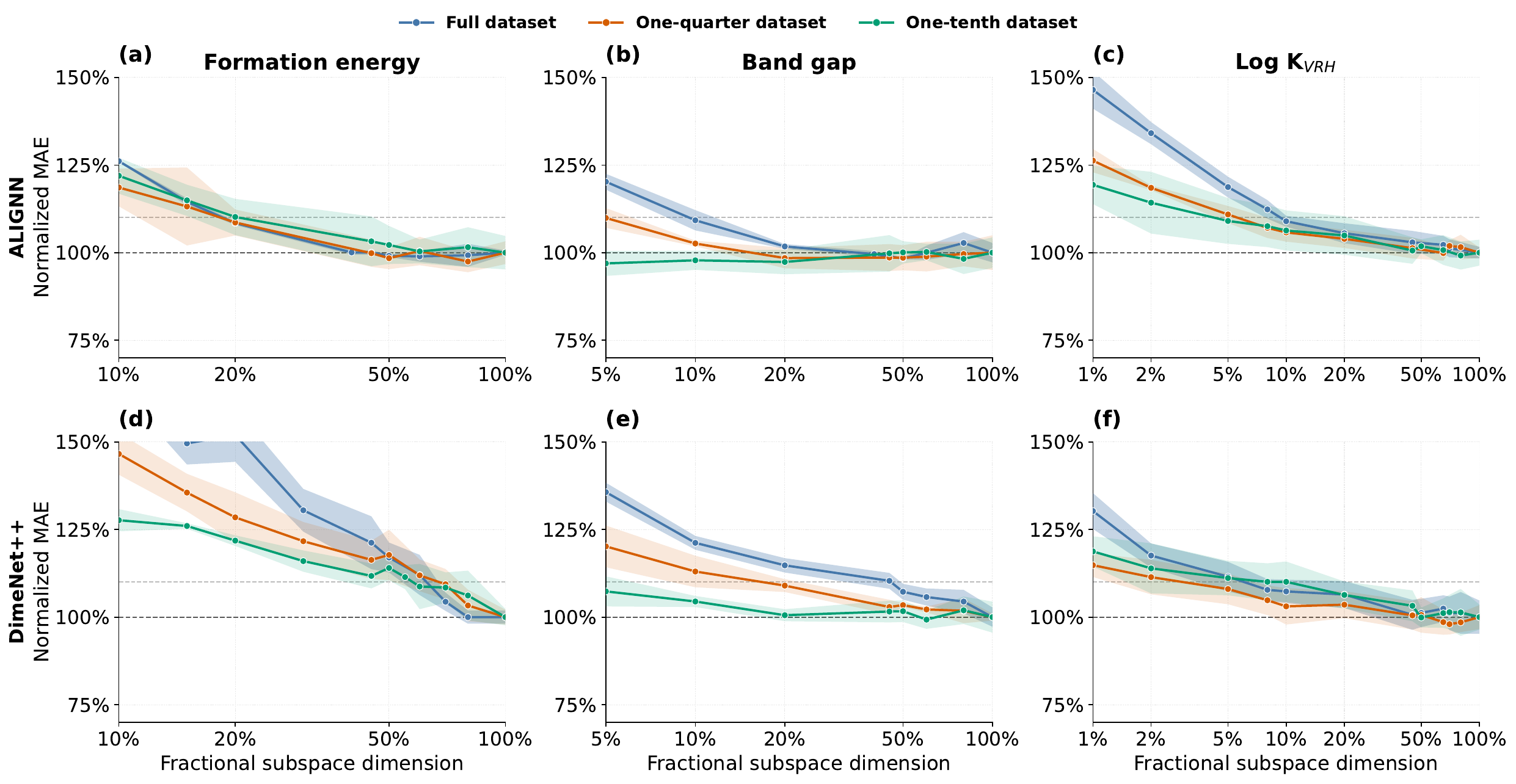}
\caption{Performance of ALIGNN and DimeNet++ on different subsets of the dataset for formation energy, band gap, and log $K_\text{VRH}$. For each subset of the dataset, 4--5 different splits were used, and each split was trained with two different model seeds. Lines and markers denote the mean across repeated runs, and shaded bands show the standard deviation under the randomization protocol described in the Methods Section. The horizontal black line marks the $d=D$ reference, and the gray dashed line marks the 10\% performance threshold used to define $d_{10}$.}
\label{Fig_quarter_formation}
\end{figure*}

To examine how dataset size affects the dimensional response, we repeated the formation-energy, band-gap, and log-$K_{\mathrm{VRH}}$ experiments using the full datasets and randomly sampled one-quarter and one-tenth subsets (Figure~\ref{Fig_quarter_formation}). For each reduced dataset size, we generated up to five independently sampled subsets, and each subset was trained using two model initializations. The model configurations and number of training epochs were held fixed across dataset sizes. For each dataset size, the dimensional-response curve is normalized by the corresponding $d=D$ performance at that same size. The curves therefore measure how readily a model recovers its own full-subspace performance as the number of training structures changes, rather than directly comparing absolute errors across dataset sizes. 

For formation-energy prediction, the normalized curves retain broadly similar shapes across the three dataset sizes. For ALIGNN, all three curves enter the 10\% tolerance region at approximately 20\% of the ambient dimension. DimeNet++ requires substantially larger fractional subspaces, with the three dataset sizes approaching the tolerance region at approximately 60--70\% of $D$. Within each architecture, however, there is no consistent monotonic ordering of the threshold with dataset size. The uncertainty bands are generally broader for the reduced formation-energy datasets, particularly for the one-tenth subsets, whereas the full-dataset response is more tightly resolved over much of the moderate- and high-dimensional regime. Together with the reduction in absolute MAE, this suggests that the main effect of increasing the formation-energy dataset is improved predictive accuracy and more reproducible training outcomes, rather than a large systematic shift in the mean fractional dimensional demand. Because the uncertainty combines variation across sampled subsets and model initializations, the present results do not identify the specific source of the increased variability at smaller dataset sizes.

Band-gap prediction shows a clearer and more systematic dependence on dataset size. For both architectures, models trained on smaller datasets recover their corresponding $d=D$ performance using smaller fractional subspaces. For ALIGNN, the full-dataset curve enters the 10\% tolerance region at approximately 10\% of the ambient dimension, whereas the one-quarter dataset is at the tolerance boundary near 5\% and the one-tenth dataset remains within the tolerance at the smallest tested dimension. The latter therefore provides only an upper bound of $d_{10}/D \leq 5\%$.

The effect is substantially larger for DimeNet++. The full-dataset model requires approximately 40--50\% of the ambient dimension to recover performance within 10\% of its $d=D$ reference, compared with approximately 20\% for the one-quarter subsets and no more than 5\% for the one-tenth subsets. Thus, increasing the amount of band-gap training data produces a progressive increase in the fractional subspace required to recover the corresponding full-subspace performance, with the effect being particularly pronounced for DimeNet++.

This ordering is consistent with the larger band-gap datasets imposing additional constraints on the learned mapping: the reduced datasets can recover their own full-subspace baselines using relatively few trainable directions, whereas recovering the improved performance obtained from the full dataset requires a larger fraction of the parameter space. The result does not, however, establish that the additional structures are chemically or structurally nonredundant. Such a conclusion would require independent measures of dataset coverage or a comparison between random and diversity-selected subsets.

Log-$K_{\mathrm{VRH}}$ shows a much weaker dependence on dataset size. For ALIGNN, the one-quarter and one-tenth subsets enter the 10\% tolerance region at approximately 5\% of the ambient dimension, whereas the full dataset requires roughly 8--10\%. For DimeNet++, the quarter dataset reaches the tolerance region at approximately 5\%, while the full and one-tenth datasets require approximately 8--10\%. The ordering is therefore not monotonic with dataset size, and the uncertainty bands overlap substantially across most of the sweep.

The log-$K_{\mathrm{VRH}}$ results consequently indicate that the normalized dimensional response is comparatively stable under dataset subsampling. Increasing the dataset size improves the attainable absolute performance, but it produces only modest and architecture-dependent changes in the fractional threshold. The present figure also does not show a clear monotonic reduction in run-to-run variability with dataset size, so any such effect would need to be established through a separate numerical summary of the variances.

Overall, the effect of dataset size on dimensional response is property dependent. Band-gap prediction shows a systematic increase in fractional dimensional demand as the training set grows, especially for DimeNet++, whereas formation energy and log $K_{\mathrm{VRH}}$ retain comparatively similar mean responses across dataset sizes. These results demonstrate that dataset growth can alter the dimensional response, but the magnitude and direction of that change depend on both the target property and the model architecture. The curves do not by themselves determine whether the additional structures contribute nonredundant chemical or structural information.

\subsection*{Dimensional demand scales with model size}
\label{sec:model_size}

We next examine whether the measured dimensional demand is robust to changes in ambient model size. This question is motivated by the model-size experiments of Li et al.~\cite{li2018intrinsicdimension}, who reported that the absolute intrinsic dimension varied little across fully connected networks whose parameter counts differed by more than an order of magnitude. To perform an analogous test for atomistic GNNs, we varied the channel width of ALIGNN and DimeNet++ while holding the prediction task, dataset, and training protocol fixed. All models were trained to predict log $K_{\mathrm{VRH}}$. 

\begin{figure*}[tb]
\centering
\includegraphics[width=\textwidth]{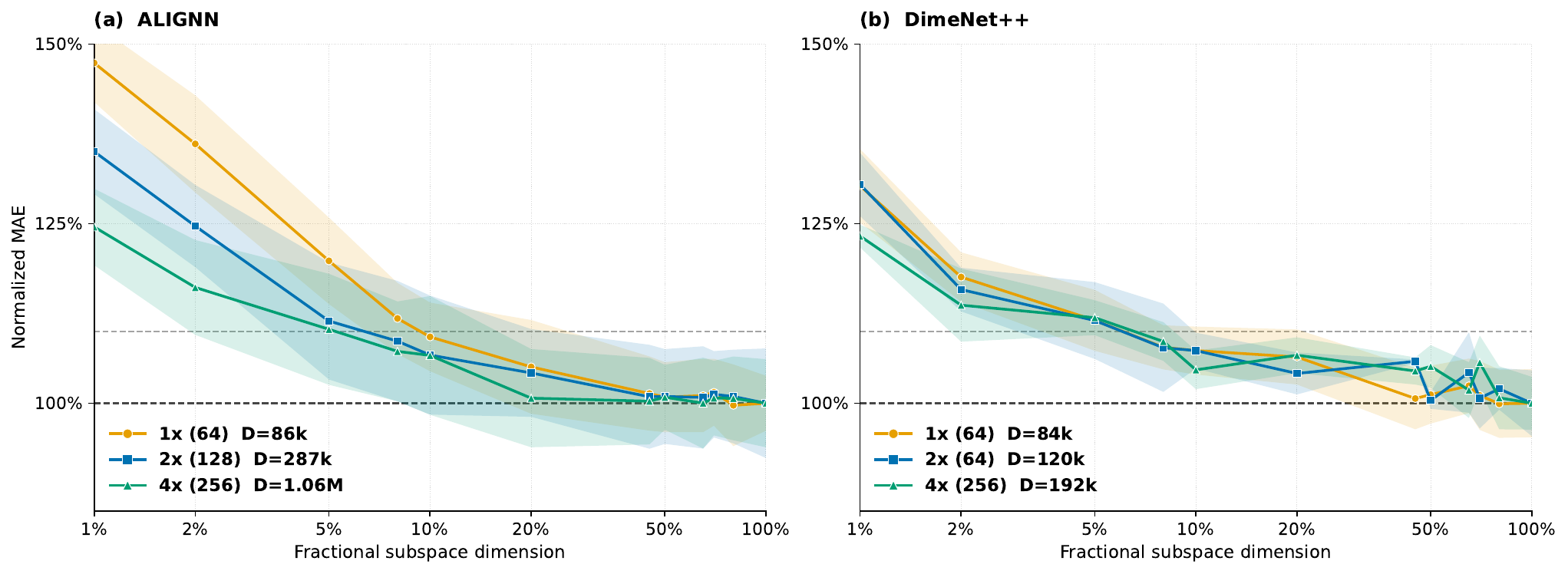}
\caption{Effect of model width on the dimensional response for log-$K_{\mathrm{VRH}}$ prediction. Each curve shows the MAE relative to the corresponding $d=D$ projected run as a function of fractional subspace dimension $d/D$. The legends report the channel width and ambient parameter dimension $D$ for each model. Lines and markers denote the mean across repeated runs, and shaded bands show the standard deviation. The horizontal black line marks the $d=D$ reference, and the gray dashed line marks the 10\% tolerance used to define $d_{10}$. For the sweep, only the convolutional segments of the layers were expanded, and other components of layers such as input and output embeddings were kept fixed.}
\label{fig_intrinsic_con_width}
\end{figure*}

For ALIGNN, increasing the channel width from 64 to 128 and 256 raises the ambient parameter dimension from approximately $86\,000$ to $287\,000$ and $1.06$ million, respectively. For DimeNet++, the corresponding configurations contain approximately $84\,000$, $120\,000$, and $192\,000$ parameters. Figure~\ref{fig_intrinsic_con_width} shows the normalized MAE as a function of the fractional subspace dimension $d/D$ for these width-scaled models.

For ALIGNN, the three normalized curves enter the 10\% tolerance region at broadly similar fractional dimensions. The wider models show somewhat less degradation in the lowest-dimensional regime, but $d_{10}/D$ remains within a comparatively narrow range as the ambient parameter count increases. Viewed only in fractional coordinates, the dimensional response therefore appears approximately robust to model width.

The conclusion changes when the threshold is expressed in absolute dimensions. Because the ALIGNN parameter count increases by approximately a factor of twelve, the similar fractional thresholds correspond to a strong increase in the number of trainable coordinates required to recover each model's $d=D$ performance. Using the thresholds extracted from the underlying sweeps, the absolute value of $d_{10}$ increases by approximately twelvefold between the smallest and largest ALIGNN configurations. Thus, the ALIGNN results are consistent with
\[
d_{10}\propto D,
\]
while the fractional requirement $d_{10}/D$ remains approximately stable.

This behavior differs from the model-size robustness reported by Li et al.~\cite{li2018intrinsicdimension}. In their experiments, the ambient parameter dimension changed by a factor of 24.1 while the absolute threshold dimension changed by only approximately 1.3--1.4 times, depending on the baseline convention. In the present ALIGNN experiment, model-size invariance is observed only after normalization by $D$; the absolute dimensional demand grows strongly with model size. The result therefore does not support the interpretation that additional ALIGNN parameters merely enlarge the set of redundant solution directions while leaving the number of required trainable degrees of freedom fixed.

The DimeNet++ curves also show similar fractional responses across the three width settings, with thresholds in the approximate range of 5--8\% of the ambient dimension. The corresponding parameter counts for DimeNet++ differ by about 130\%, from approximately $84\,000$ to $192\,000$. The results point towards a similar trend as that of ALIGNN; the dimensional demand is proportional to model size. However, since the growth in parameter count is not as significant as that for ALIGNN, we therefore interpret the DimeNet++ result only as evidence of local stability of the fractional response over the tested width range.

Because each curve is referenced to its own $d=D$ performance, these experiments measure the number of dimensions required to recover each model's own full-subspace baseline. Differences in the attainable $d=D$ MAE may therefore contribute to the observed scaling and should be considered alongside $d_{10}$ and $d_{10}/D$. Moreover, the ALIGNN conclusion is based on three model sizes for a single prediction task. We consequently interpret the result as a breakdown of absolute model-size invariance under the present protocol, rather than as a universal scaling law for atomistic GNNs.

\section*{Discussion}\label{sec12}

This study shows that final predictive accuracy and trainable-degree dependence characterize different aspects of a materials-learning system. A final MAE or ROC-AUC value reports the endpoint reached by a model. A random-subspace sweep shows how that endpoint is recovered as independent trainable directions are progressively restored. The resulting curve is a response measurement that shows how strongly the achieved performance depends on access to the training space and how reproducibly that dependence appears across repeated runs.

This perspective separates effects that are mixed together in a table of final errors. A low final MAE can come from a solution that is recovered with relatively few trainable directions, or from one that requires broad access to the optimization space. Strong degradation at small subspace dimension indicates that the model's full-subspace performance depends on many coordinated parameter directions, even when the endpoint accuracy is good. The phonon results illustrate this distinction: the architecture with the lowest full-subspace error is not the one with the weakest relative degradation under dimensional restriction. Random-subspace analysis therefore adds information beyond model ranking by showing whether an accuracy advantage remains robust when trainable freedom is severely reduced.

The dataset-size experiments suggest that recovery curves can also distinguish different effects of adding data. Increasing a dataset can improve endpoint accuracy by stabilizing the same learned mapping, or it can introduce additional constraints that require more trainable degrees of freedom to satisfy. In the band-gap experiments, larger datasets require larger fractional subspaces to recover their own improved full-subspace performance, consistent with the second case. Formation energy and log $K_{\mathrm{VRH}}$ show weaker shifts in the mean recovery curve, suggesting a regime in which additional data improves accuracy or stability without strongly changing model-relative dimensional demand. This does not establish chemical redundancy or diversity by itself, but it gives a concrete way to test those hypotheses in future studies by comparing random subsets with diversity-selected or redundancy-enriched subsets.

The model-width experiment gives a related lesson for interpreting intrinsic-dimension-style measurements. In the ALIGNN width sweep, the fractional threshold $d_{10}/D$ remains approximately stable while the absolute threshold $d_{10}$ grows strongly with the ambient parameter dimension. Thus, the question ``how many dimensions are needed'' has two distinct answers. The absolute threshold measures how many trainable coordinates are required in the chosen parameterization, while the fractional threshold measures what fraction of that parameterization is required. Both quantities carry information, and both should be reported when model size changes. This result makes the measured dimensional demand a property of the model, dataset, task, and training protocol together, rather than a task-only descriptor.

These findings also clarify how the method can be used. Random-subspace analysis is most useful as a targeted diagnostic when final accuracy leaves an important ambiguity: two models have similar errors but different apparent robustness, a more accurate model appears unstable, dataset growth improves performance in some tasks but not others, or model scaling changes the interpretation of intrinsic dimension. In such cases, the recovery curve can reveal whether performance depends strongly on access to the full training space, whether degradation occurs gradually or sharply as dimensions are removed, and whether the response is stable across randomizations. This makes the method suitable for focused model-comparison, dataset-analysis, and ablation studies.

Several limitations define the scope of the present results. The main comparisons use compact one-layer configurations selected for repeated sweeps rather than state-of-the-art accuracy. The scalar threshold $d_{10}$ summarizes a full recovery curve and can be uncertain for nonmonotonic responses or coarse dimension grids. The uncertainty bands combine several sources of randomness, and each curve is normalized to its own $d=D$ reference, so dimensional efficiency means recovery of model-relative performance rather than achievement of a common absolute error. The classification threshold based on raw ROC-AUC is also not quantitatively equivalent to the regression threshold based on relative MAE.

Overall, random-subspace intrinsic-dimension analysis adds trainable-degree dependence as a second axis for understanding materials GNNs. It asks how much independent optimization freedom a model--dataset--property system needs to recover its own reference performance, and how stable that requirement is. This response-based view connects architecture, data, optimization, and target property in a single measurement. Future work can use it to separate sources of variability, relate dimensional demand to chemical and structural diversity, test out-of-distribution settings, and examine whether the same qualitative signatures persist in larger materials GNNs.

\section*{Methods}\label{sec2}

\subsection*{Random-subspace training}

Let $\boldsymbol{\theta}\in\mathbb{R}^{D}$ denote the complete trainable parameter vector of a model. Instead of optimizing all $D$ parameters directly, we restrict training to a randomly oriented affine subspace of dimension $d\leq D$:
\begin{equation}
\boldsymbol{\theta}
=
\boldsymbol{\theta}_{0}
+
\mathbf{A}\boldsymbol{\phi},
\label{eq1}
\end{equation}
where $\boldsymbol{\theta}_{0}$ is a fixed model initialization, $\mathbf{A}\in\mathbb{R}^{D\times d}$ is a fixed random projection, and $\boldsymbol{\phi}\in\mathbb{R}^{d\times 1}$ is the trainable coordinate vector. Training therefore solves
\begin{equation}
\min_{\boldsymbol{\phi}}
\mathcal{L}
\left(
\boldsymbol{\theta}_{0}
+
\mathbf{A}\boldsymbol{\phi}
\right),
\label{eq2}
\end{equation}
with gradients
\begin{equation}
\nabla_{\boldsymbol{\phi}}\mathcal{L}
=
\mathbf{A}^{\top}
\nabla_{\boldsymbol{\theta}}\mathcal{L}.
\end{equation}

Repeating the training procedure over increasing values of $d$ produces the dimensional-response curve. The method measures how much randomly oriented trainable freedom is required to recover a specified level of predictive performance; it does not identify a sparse subset of model parameters or reduce the size of the deployed architecture.





\subsection*{Models, tasks, and dataset-size experiments}

We study three atomistic GNN architectures: CGCNN, ALIGNN, and DimeNet++. For the principal cross-architecture comparison, compact configurations were selected with approximately matched parameter counts: 86,617 parameters for CGCNN, 85,963 for ALIGNN, and 83,685 for DimeNet++. These configurations were chosen to make repeated dimensional sweeps computationally feasible rather than to reproduce the largest or most accurate versions of the original models.

The six prediction tasks are formation energy per atom, band gap, metallicity classification, log $K_{\mathrm{VRH}}$, refractive index, and phonon frequency. Formation energy and band gap are taken from MP-23, while the remaining tasks use the corresponding Matbench datasets. Regression tasks are evaluated using MAE and metallicity classification using ROC-AUC. For all tasks, the data are randomly split into 80\% training, 10\% validation, and 10\% test sets using fixed random seeds.

To study dataset-size effects, the formation-energy, band-gap, and log-$K_{\mathrm{VRH}}$ experiments are repeated using the full training data and randomly sampled one-quarter and one-tenth subsets. Each dataset size is normalized by its own $d=D$ result, so the comparison measures how readily a model recovers the full-subspace performance available at that dataset size. Up to five sampled subsets are used at each reduced size, with two model initializations per subset.

\subsection*{Projection, optimization, and uncertainty}

The main experiments use Fastfood structured projections, which avoid explicitly storing the full $D\times d$ projection matrix. Projection sensitivity is evaluated using dense, Fastfood, dense-plus-orthonormalized, and Fastfood-plus-orthonormalized constructions. For the orthonormalized variants,
\begin{equation}
\mathbf{A}^{\top}\mathbf{A}
=
\mathbf{I},
\end{equation}
which removes unequal scaling and correlation among the subspace coordinates while preserving the orientation of the sampled subspace.

As the main sweeps use Fastfood structured projections, we performed a representative projection-construction control using DimeNet++ on log-$K_{\mathrm{VRH}}$. Dense, Fastfood, dense-orthonormalized, and Fastfood-orthonormalized projections produced the same broad recovery behavior, with all mean curves reaching the 10\% tolerance by $d/D=10\%$ (Supplementary Figure~S1). This supports the use of Fastfood for the large experimental sweep, while leaving open possible projection-construction effects in other model--task combinations.

All models are trained using AdamW with a one-cycle learning-rate schedule. The model configuration, optimizer, scheduler, and number of epochs are held fixed within each comparison. Exact architecture hyperparameters, learning rates, batch sizes, epoch counts, and projection settings are provided in the Supplementary Information (Table~S1-S3).

The reported curves are averages over repeated runs, and the shaded regions denote one standard deviation. Depending on the experiment, repeated runs vary sampled data subsets, model initialization, and projection orientation. The uncertainty bands therefore quantify overall run-to-run sensitivity under the corresponding randomization protocol and are not attributed to a single source unless that source is varied independently.

\section*{Data availability}
The formation energy and band gap datasets are available on Zenodo at: \url{https://doi.org/10.5281/zenodo.21871852}. The bulk modulus, refractive index,
phonon-frequency, and is-metal classification datasets were downloaded from Matbench at: \url{https://matbench.materialsproject.org/}.

\section*{Code availability}
The code used in this work is available on GitHub at: \url{https://github.com/shehrozashoaib/Intrinsic_Dimensionality_ML_Materials}.

\section*{Acknowledgements}

The research was supported by funding from King Abdullah University of Science and Technology (KAUST). For computer time, this research used Shaheen III and Ibex managed by the Supercomputing Core Laboratory at King Abdullah University of Science and Technology (KAUST) in Thuwal, Saudi Arabia.

\section*{Author Contributions}
K.L. conceived and supervised the project. S.A.S. performed the experiments and analyzed the results. S.A.S. and K.L. wrote and reviewed the manuscript. 

\section*{Competing interests}
The authors declare no competing interests.

\bibliography{sn-bibl}

\end{document}